%% file: BottomUp.tex
\documentclass[letterpaper, 10pt, conference]{ieeeconf}

\IEEEoverridecommandlockouts                              
\usepackage{graphicx}
\usepackage[dvipsnames]{xcolor}
\usepackage{url}
\usepackage{amsmath} 
\usepackage{amssymb}  
\usepackage{amsfonts}%
\usepackage[mathscr]{eucal}

\let\labelindent\relax
\usepackage{enumitem}

\makeatletter
\let\NAT@parse\undefined
\makeatother
\usepackage[plainpages = false, colorlinks=true, linkcolor=Blue, citecolor = MidnightBlue, urlcolor = blue, filecolor=black, pagebackref=false, hypertexnames=false,  pdfpagelabels ]{hyperref}

\newtheorem{thm}{Theorem}

\newtheorem{rem}[thm]{Remark}

\newcommand{\trasp}{\ensuremath{^{\intercal}}}
\newcommand{\bfx}{\ensuremath{\mathbf{x}}}
\newcommand{\bfu}{\ensuremath{\mathbf{u}}}

\newcommand{\setS}{\ensuremath{\mathcal{S}}}
\newcommand{\setP}{\ensuremath{\mathcal{P}}}

\newcommand{\setN}{\ensuremath{\mathcal{N}}}

\title{\LARGE \bf Coalitional control: a bottom-up approach*}
\author{Filiberto Fele, Jos\'e M.~Maestre and Eduardo F.~Camacho
\thanks{*Financial support by the FP7-ICT project DYMASOS (ref. 611281), and Junta de Andaluc\'{i}a  (ref. P11-TEP-8129) is gratefully acknowledged.}
\thanks{The authors are with Departamento de Ingenier\'{i}a de Sistemas y Autom\'{a}tica, ETSI Universidad de Sevilla, 41092 Seville, Spain
        {\tt\small ffele@us.es}, {\tt\small pepemaestre@us.es}, {\tt\small efcamacho@us.es}}%
\thanks{\copyright 2015 AACC. The final version of the article is available at \url{https://doi.org/10.1109/ACC.2015.7171966} (please cite as \cite{FeleEtAl2015ACC}).}
}
\begin{document}
\maketitle
\thispagestyle{empty}
\pagestyle{empty}
%
\begin{abstract}
The recent major developments in information technologies have opened interesting possibilities for the effective management of multi-agent systems. In many cases, the important role of central control nodes can now be undertaken by several controllers in a distributed topology that suits better the structure of the system. This opens as well the possibility to promote cooperation between control agents in competitive environments, establishing links between controllers in order to adapt the exchange of critical information to the degree of subsystems' interactions. In this paper a bottom-up approach to \emph{coalitional control} is presented, where the structure of each agent's model predictive controller is adapted to the time-variant coupling conditions, promoting the formation of coalitions --- clusters of control agents where communication is essential to ensure the cooperation --- whenever it can bring benefit to the overall system performance.
\end{abstract}

\input{intro}
\input{prob_stat}
\input{coal_alg}
\input{sim}
\input{conclusions}


\bibliographystyle{IEEEtran}
\bibliography{biblio_tesi}
\end{document}

%% file: intro.tex
\section{Introduction}
Whether dealing with a single plant composed of a structured set of subsystems (e.g., chemical processing plant) or a system arising from the interaction of different entities (e.g., microgrid), it is generally possible to identify a set of coupled control subproblems that jointly configure the global one. Indeed, there are cases in which centralized strategies do not exploit the structure of the system, leading to oversized computational and communicational requirements. Within the model predictive control (MPC) framework, a consistent effort has been put in the last decade for the development of non-centralized control schemes for large-scale systems, further expanding the frontiers of MPC applications~\cite{SCA09JPC,NegenbornMaestreCSM2014,StewartEtAl2010,CAM02IEEECSM}. The common objective of such approaches is to achieve the scalability and robustness inherent in a distributed implementation, maintaining a global optimal performance comparable to that expected through the use of a centralized controller.\par
The developments in data acquisition technologies (wireless networks, smart sensors) and in database management (cloud computing) has yielded means of sharing measures and other information of a large-scale system in an efficient and flexible way~\cite{VadigepalliDoyle2003} and --- thanks to the wide diffusion of smart devices --- enhanced the possibilities offered by human-in-the-loop control systems~\cite{MaestreEtAl14CDC}. Several examples of the advances introduced by these new technologies on infrastructure systems can be identified, e.g., on traffic, water, and electricity networks~\cite{NEGENBORN10BOOK}. Such improvement in the computational and communicational capabilities provided to local (mobile) control devices constitutes an additional impulse towards a new approach to distributed control problems: one where the cooperation between networked controllers is actively fostered and adapted in real-time to the state of the system.\par
The design of a distributed control strategy involves a tradeoff between performance and complexity of implementation. Whenever possible it is desirable to formulate control laws based exclusively on local information, in order to reduce the computational requirements and the communication overhead~\cite{Siljak1991,ZecevicSiljak2010}. In general, though, the denser the interaction among different parts of a system, the more intense the communication required between the control agents --- the extreme case corresponding to a distributed solution of the centralized control problem~\cite{RawlingsStewart2008}. Nonetheless, in several cases the variables of a system can be grouped to highlight weakly coupled blocks, often revealing a natural topology. Within each block (usually designated as \emph{neighborhood}) dynamic interactions propagate quickly, affecting the rest of the system on a longer time scale~\cite{StewartEtAl2010_ACC}. Therefore, it is an interesting challenge to identify online the degree of interaction, in order to consequently adjust the structure of the controller; such flexibility would also grant the possibility of accommodating the computational and communicational requirements in real-time.\par
All these issues and challenges constitute the rationale leading to \emph{coalitional control}, where the control strategy is adapted to the varying coupling conditions between the agents, promoting the formation of coalitions --- clusters of control agents where communication is essential to ensure the cooperation --- whenever it can bring benefit to the overall system performance~\cite{FeleJPC,MAESTRE13OCAM,TroddenRichards2009,JilgStur2013_IFAC}.\par
The rest of this document is organized as follows: in Section~\ref{sec_sysmodel} the control problem is stated; the proposed bottom-up approach for coalitional control is described in~\ref{sec_coal_schemes}; then, results from simulations carried out on a flow network model are presented in Section~\ref{sec_sim}. Finally, Section~\ref{sec_concl} provides conclusions and an outlook on future work.

%% file: prob_stat.tex
\section{Problem statement}
\label{sec_sysmodel}
Consider a set $\setS = \{1,\ldots,M\}$ of dynamically coupled discrete-time linear processes, each modeled by the following state-space equation:
\begin{equation}\label{eq_ss}
x_i(k+1)=A_i x_i(k)+ \sum\limits_{j \in \setN_{i}}B_{ij} u_{ij}(k) + \sum\limits_{j \in \setN_{i}}B_{ji} u_{ji}(k),
\end{equation}
where $x_i\in\mathbb{R}^n$ is the state of the $i$th subsystem, $u_{ij}\in\mathbb{R}^m$ is the control action applied by subsystem $j$ on subsystem $i$ (and viceversa for $u_{ji}$). 
$\setN_i\subset\setS$ is the set of indices that identify the subsystems coupled through the input to subsystem $i$. Real large-scale systems that show analogies with the above formulation are, e.g., drinking water networks~\cite{Ocampo2012HierarcMPC}, which are composed of interconnected water tanks, irrigation canals, which have been modeled by integrator-delay models~\cite{FeleJPC,Negenborn:08c}, and supply chains~\cite{JMM09CDCb,JMM11JPC}. In general, such systems show a natural complex nonlinear behavior: linearizations such as~\eqref{eq_ss} are employed at the control layer under given assumptions about the operating conditions of the system (see, e.g.,~\cite{Schuurmans1997}).\par
In order to simplify the notation, without loss of generality the coupling is considered here to be ``symmetric''. 
In other words, $\setN_i$ identifies all subsystems whose control actions influence the trajectory of $x_i$, as well as all subsystems whose state is affected by the control actions of subsystem $i$.
Finally, $A_i\in\mathbb{R}^{n\times n}$, $B_{ij}\in\mathbb{R}^{n\times m}$ and $B_{ji}\in\mathbb{R}^{n\times m}$ are constant state transition matrices. Constant delays can be modeled by considering an augmented state vector.\par
The performance of each local controller $i\in\setS$ is measured through a stage cost defined as:
\begin{multline}\label{eq_ind_cost}
\ell_i(k)=(x_i(k) -\bar x_i )\trasp Q_i (x_i(k)-\bar x_i) \\+ (u_i(k)-\bar u_i)\trasp R_i (u_i(k)-\bar u_i),
\end{multline}
\noindent where $x_i$ and $u_i\triangleq \{u_{ji}\}_{j\in\setN_i}$ are respectively the state and input vectors of subsystem $i$. The matrices $Q_i$ and $R_i$ are weighting matrices that penalize the deviation of state and input from their reference.\par
We consider the scenario in which the agents in the system act in order to minimize their stage cost~\eqref{eq_ind_cost}. In the basic decentralized architecture considered here, each subsystem is governed by a local MPC controller, whose input actions are obtained as the solution of the following optimization problem: 
\begin{equation}
\min\limits_{\mathbf{u}_i} \sum\limits_{t=k}^{k+N_p} \ell_i(t)
\label{eq_mpc_prob}
\end{equation}
s.t.
\begin{subequations}
\begin{align}\label{eq_mpc_prob_constr}
& x_i(t+1) = A_{i} x_i(t) + \sum\limits_{j \in \setN_i} B_{ji} u_{ji}(t)\\ 
\label{eq_mpc_prob_constr2}
& u_i(t) \in \mathcal{U}_i,\quad \forall t \in \left\{k, \ldots,k + N_p - 1\right\}
\end{align}
\end{subequations}
where the optimization variable $\mathbf{u}_i \triangleq [u_i(k),\ldots, u_i(k+N_p-1)]$
is a column vector composed of the sequence of control actions along the prediction horizon of length $N_p$. At time $k$ the first element $\mathbf{u}_i^{\ast}(0)\triangleq u_i(k)$ of the minimizing sequence is applied to the system, and the problem is solved again at subsequent time instants in a receding horizon fashion~\cite{EFC1,RawlingsLIB09}.

%% file: coal_alg.tex
\section{Coalitional control schemes}
\label{sec_coal_schemes}
This work aims to explore the enhancement provided by a dynamic management of a network infrastructure which interconnects the control agents (following the line of~\cite{MAESTRE13OCAM,FeleJPC,TroddenRichards2009}). Communication allows to establish cooperation between given clusters of agents (referred to as \emph{coalitions}), and is managed by trading off control performance for savings on communication costs. As a result, the proposed coalitional control algorithms promote an increase of the degree of reciprocal coordination among agents whenever the dynamic interaction between their relative subsystems is most critical. Ad hoc bottom-up negotiation criteria for the formation of new coalitions will be introduced next. First, the Coalitional Cooperative (COO) criterion is presented, and then individual rationality issue is considered by the Coalitional Individually Rational (CIR) criterion. The splitting process is not considered in this paper: instead, it is assumed that any coalition has a certain lifetime, after which its members are restored to the decentralized configuration.\par
In the remainder, the term \emph{player} may refer to a single control agent or, if it is the case, to a group of agents which previously joined into a coalition and act as a unique entity.\par
At given time intervals (in general multiple of the sample time), all the players will participate in pairs into a noniterative bargaining process, whose outcome will decide the generation of new coalitions. Any new coalition will be the product of the union of two players, and thus of all agents they involve. The basic criterion for the bargaining is an increase of benefit for both players, measured in terms of control performance optimality; moreover, costs required for the cooperation to take place are considered. Two different indices (designated as $(a)$ and $(b)$) to express cooperation costs are proposed and evaluated in the simulations.\par
To keep notation simple, subscripts $\{1,2\}$ will be used in the remainder to designate the players. Furthermore, subscript $12$ will refer to the merger. The subsystems involved in either part of a given bargaining process are identified by the sets $\setP_1\subset\setS$ and $\setP_2\subset\setS$. As previously stated, the coalitions of agents corresponding to these sets constitute the two players. Notice that $\mathcal{P}_1 \cap \mathcal{P}_2 = \emptyset$.\par
Then, the cooperation costs indices are defined as:
\begin{enumerate}[labelindent=\parindent,leftmargin=*,itemsep=0.5ex,label=$(\alph*)$]
	\item $\chi_{12} = f_a(|\setP_1|+|\setP_2|)$
	\item $\chi_{12} = f_b(n_{l1}+n_{l2}+1)$
\end{enumerate}
where $|\cdot|$ stands for the cardinality of a set. With index $(a)$, additional costs required for the coordination of large coalitions due to, e.g., increased bandwidth for information exchange, longer distance communications, increased computational complexity, are bundled as a function of the total number of agents involved in the bargaining process. Index $(b)$ expresses the cost of use of the data links enabled in order to establish communication between every member of the coalition. In particular, following the standard definition of \emph{connectedness} in graph theory~\cite{MSAVDN01}, it is assumed here that agents within a coalition can communicate when there exists a path of enabled links between them (e.g., multihop communication). Thus, $n_{l1}$ ($n_{l2}$) is the number of active links allowing agents in $\setP_1$ ($\setP_2$) to communicate: one further link is considered for the generation of the merger $\setP_1\cup\setP_2$, to connect the respective underlying communication graphs.\par
Following the previous definition of \emph{player}, the states and inputs of every subsystem constituting the coalitions that take part in the bargaining process are gathered into the player's state and input vectors, defined as:
\begin{align*}
x_{p1} & \triangleq\{x_i\},\; & u_{p1} & \triangleq\{u_i\},\quad & \forall i\in\mathcal{P}_1 &\subset\setS \\
x_{p2} & \triangleq\{x_j\},\; & u_{p2} & \triangleq\{u_j\},\quad & \forall j\in\mathcal{P}_2 &\subset\setS .
\end{align*} 
Also, the merger state and input vectors are composed according to
\begin{equation*}
\xi\trasp = [x_{p1}\trasp\; x_{p2}\trasp],\quad \nu\trasp = [u_{p1}\trasp\; u_{p2}\trasp].
\end{equation*}
Next, the bargaining criteria for the COO and the CIR algorithms are detailed.
\subsection{Coalitional Cooperative algorithm}
\label{sec_COO}
The joint performance index~\eqref{eq_cost_merge} for the merger is essentially composed by two terms: the first is the aggregate compound MPC cost over the horizon, while the second corresponds to a given cooperation cost index. 
\begin{equation}
J_{12} = \sum\limits_{t=k}^{k+N_p} \ell_{12}(t) + \chi_{12}.
\label{eq_cost_merge}
\end{equation}
The compound stage cost $\ell_{12}$
is formulated as
\begin{multline}\label{eq_compound_cost}
\ell_{12} = (\xi(k) -\bar{\xi} )\trasp Q_{12} (\xi(k)-\bar{\xi}) \\+ (\nu(k)-\bar{\nu})\trasp R_{12} (\nu(k)-\bar{\nu}),
\end{multline}
where
\begin{equation*}
Q_{12} = \left[\begin{array}{cc}
	Q_{1} & 0 \\
	0 & Q_{2}
\end{array}\right],\quad
 R_{12} = \left[\begin{array}{cc}
	R_{1} & 0 \\
	0 & R_{2}
\end{array}\right];
\end{equation*}
the individual player's weights are block diagonal matrices such that $Q_{1} = diag(\{Q_i\}_{i\in\setP_1})$, $R_{1} = diag(\{R_i\}_{i\in\setP_1})$, and analogously for player 2.
The stage cost associated to each single player in the case that the merger is not approved, i.e., the players act without express coordination, is defined as
\begin{multline}\label{eq_cost_player}
\ell_{i} = (x_{pi}(k) -\bar{x}_{pi} )\trasp Q_{i} (x_{pi}(k)-\bar{x}_{pi}) \\+ (u_{pi}(k)-\bar{u}_{pi})\trasp R_{i} (u_{pi}(k)-\bar{u}_{pi}),\quad i\in\{1,2\},
\end{multline}
where $Q_i$ and $R_i$ are the block diagonal matrices defined above. The corresponding player's performance index $J_i$ along the MPC prediction horizon is
\begin{equation}
J_{i} = \sum\limits_{t=k}^{k+N_p} \ell_{i}(t) + \chi_i,\quad i\in\{1,2\},
\label{eq_cost_player2}
\end{equation}
with the cooperation costs $\chi_i$ involving only player $i$ internal communication.\par
Finally, MPC problem~\eqref{eq_mpc_merger} is solved for $i\in\{1,2,12\}$, i.e., the three cost functions resulting as the outcome of agreement or disagreement on the formation of the new coalition between the two players:
\begin{equation}
J_i^{\ast} = \chi_i + \min\limits_{\mathbf{u}_i} \sum\limits_{t=k}^{k+N_p} \ell_i(t)
\label{eq_mpc_merger}
\end{equation}
The constraints for problem~\eqref{eq_mpc_merger} solved for $i \in\{1,2\}$ are:
\begin{subequations}
\begin{align}\label{eq_mpc_player_constr}
& x_{pi}(t+1) = A_{pi} x_{pi}(t) + B_{pi} u_{pi}(t)\\ 
\label{eq_mpc_player_constr2}
& u_{pi}(t) \in \bigotimes\limits_{j\in\setP_i}\mathcal{U}_j,\quad \forall t \in \left\{k, \ldots,k + N_p - 1\right\}
\end{align}
\end{subequations}
where $A_{p1} = diag(\{A_i\}_{i\in\setP_1})$ and $B_{p1} = \left[B^{(ij)}\right]$, with
\begin{align*}
 B^{(ii)} & = \left[B_{{s_1}i}\cdots B_{{s_N}i}\right],\quad & s & \in\setN_i,\, N=|\setN_i|\\
 B^{(ij)} & = \left[\cdots B_{ij}\cdots\right],\quad & j &\in\setP_1,\,j\in\setP_1\setminus \{i\};
\end{align*}
notice that $B_{ij}\neq \mathbf{0}$ if and only if $j\in\setP_1\cap\setN_i$. Matrices for player 2 are composed likewise.\par
Constraints on problem~\eqref{eq_mpc_merger} solved for the merger are
\begin{subequations}
\begin{align}\label{eq_mpc_merger_constr}
& \xi(t+1) = \Xi \xi(t) + \Upsilon \nu(t)\\ 
\label{eq_mpc_merger_constr2}
& \nu(t) \in \bigotimes\limits_{j\in\setP_1\cup\setP_2}\mathcal{U}_j,\quad \forall t \in \left\{k, \ldots,k + N_p - 1\right\},
\end{align}
\end{subequations}
The state transition matrices in~\eqref{eq_mpc_merger_constr} are composed as
\begin{equation*}
\Xi = \left[\begin{array}{cc}
	A_{p1} & 0 \\
	0 & A_{p2}
\end{array}\right],\quad
\Upsilon =  \left[\begin{array}{cc}
	B_{p1} & \Upsilon^{(12)} \\
	\Upsilon^{(21)} & B_{p2}
\end{array}\right],
\end{equation*}
where $\Upsilon^{(12)} = \left[B^{(ij)}\right]$,
\begin{equation*}
 B^{(ij)} = \left[\cdots B_{ij}\cdots\right],\quad  i \in\setP_1,\,j\in\setP_2,
\end{equation*}
with $B_{ij}\neq \mathbf{0}$ if and only if $j\in\setP_2\cap\setN_i$. An analogous definition holds for $\Upsilon^{(21)}$.\par
Finally, the coalition $\setP_1\cup\setP_2$ is formed if and only if the following condition is verified:
\begin{equation}
J_{12}^{\ast} \leq J_{1}^{\ast} + J_{2}^{\ast},
\label{eq_COO_OK}
\end{equation}
where the superscript $\ast$ designates the value of~\eqref{eq_cost_merge} and~\eqref{eq_cost_player2} corresponding to the minimizing input sequence obtained by the solution of~\eqref{eq_mpc_merger}. Notice that for $i\in\{1,2\}$, the problem is solved without considering the effect of unknown inputs from other subsystems (as in~\eqref{eq_mpc_prob_constr}).
\subsection{Coalitional Individually Rational algorithm}
The condition~\eqref{eq_COO_OK} expresses the enhancement of the joint performance achieved with cooperation, w.r.t. that obtained through separate control sequence optimization. However, the derived surplus is not straightforwardly allocated to each of the players. Therefore, an additional step has to be performed in order to redistribute the coalition cost~\eqref{eq_cost_merge} over its members, such that they are offered better conditions than those they could achieve on their own (i.e., \emph{individual rationality} is satisfied). Following such allocation, the players will then have to adjust their share of the costs of the coalition by means of proper side payments.\par
Here, such allocation is computed through the Shapley value~\cite{MyersonLibroGT} for a two-player cooperative game:
\begin{multline}
\phi_i(J_{12}) = \frac{1}{2} v(\setP_i) + \frac{1}{2} \left[v(\setP_1 \cup \setP_2) - v(\setP_j)\right] \\ \triangleq \frac{1}{2} J_{i} + \frac{1}{2} (J_{12}-J_{j}),\quad j\in\{1,2\}\setminus\{i\}
\label{eq_Shapley}
\end{multline}
According to this definition, the payoffs computed through~\eqref{eq_Shapley} represent the contributions of each player to the total cost index $J_{12}$.
If the joint performance~\eqref{eq_cost_merge} is such to guarantee the following condition: 
\begin{equation}
\phi_1 \leq J_{1} \; \wedge \; \phi_2 \leq J_{2},\label{eq_cond_p1p2}\\
\end{equation}
i.e., the outcome of the merger is (individually) advantageous for both players, then $\setP_1 \cap \setP_2$ is formed by enabling communication between all the agents involved. Notice that necessary and sufficient condition for the compliance of~\ref{eq_cond_p1p2} is that $J_{12}\leq J_{1} + J_{2}$.\par
\begin{rem}
Whenever a given set of agents participate into a coalition, they act as a single entity. In this work, the bargaining is performed over the cost of the player as a whole and not over the cost of its individual components. Thus, it is assumed that an agreement that is beneficial for the entire coalition will be beneficial for each of its members too. However, this approximation of the problem avoids combinatorial explosion of the possible configurations that would arise otherwise (also when dealing with a fairly low number of agents).
\end{rem}
%

%% file: sim.tex
\section{Simulations}
\label{sec_sim}
\begin{figure*}[tb]
  \includegraphics[width=\textwidth, trim = 0mm 7mm 0mm 0mm, clip=true]{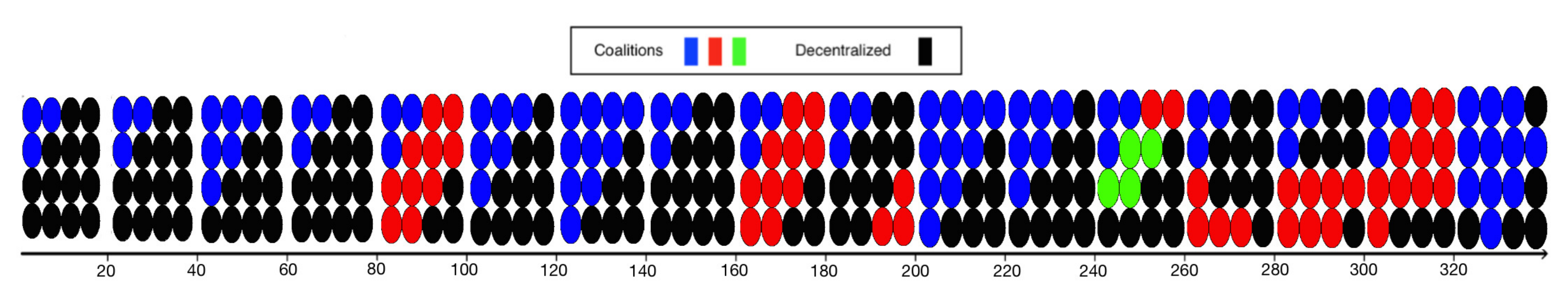}
  \caption{Evolution of the coalitions produced by the COO algorithm, with cooperation costs expressed as $(b)$. The cooperation between subsystems is promoted in order to improve their MPC performance index and, as a consequence, the overall system is driven towards its setpoint. As the cost decreases, the cooperation rate is reduced.}
  \label{fig:ejemplo}
\end{figure*}

A simple example of a storage network is used to illustrate the coalitional control framework proposed in this paper. It consists of a set $\setS = \{1,\ldots,M\}$ of integrators that are coupled through the inputs. The integrators are arranged into a 4 by 4 matrix (as schematized in Fig.~\ref{fig:ejemplo}), and each (internal) node $i\in\setS$ is connected to its four neighbors (up-down-left-right). Each subsystem behaves according to the following model:
\begin{equation}\label{eq_model}
x_i(k+1)=x_i(k)+T_s (\sum\limits_{j \in \setN_i}u_{ij}(k) - \sum\limits_{j \in \setN_i}u_{ji}(k)),
\end{equation}
where $x_i$ is the state of subsystem $i$, representing the storage level, $\setN_i$ is the set of subsystems coupled to subsystem $i$ as defined in Section~\ref{sec_sysmodel}, $u_{ij}$ and $u_{ji}$ are the inflows and outflows --- to and from subsystem $j\in\setN_i$ respectively --- and $T_s$ is the sampling time. In this particular case, it holds that if $j\in\setN_i$, then $i\in\setN_j$.\par
The goal is to regulate the state of all systems $i\in\setS$ to the target setpoint $\bar{x}_i=0.5$. The initial value for the state is $x_i(0) = 0.25$, $\forall i\in\setS$. The following constraints are imposed on control variables:
\begin{equation*}
u_i \triangleq u_{ji} \geq \mathbf{0},\quad \forall i\in\setS,\, j\in\setN_i,
\end{equation*}
i.e., the outflows towards the neighbors are constrained to be positive, so that each subsystem is only able to manipulate the decrease rate of its level. Inflows $u_{ij}$ are manipulated by the neighboring agents $j\in\setN_i$ and, from node $i$'s point of view, constitute an unmodeled disturbance in~\eqref{eq_mpc_prob_constr}. The only exception is represented by the top-left node in the 4 by 4 matrix, which is in charge of pumping water from the source into the rest of the system. Likewise, the bottom-right node acts as a sink. Given such setting, the employ of a coordination scheme in order to achieve the goal is motivated.\par
In order to provide a bound on the optimal overall performance, the two coalitional control algorithms presented in Section~\ref{sec_coal_schemes} have been compared in the simulations with centralized and decentralized MPC configurations:
\begin{itemize}
\item Centralized control (CEN): a centralized MPC controller drives the overall system to the setpoint. The performance index to be minimized in~\eqref{eq_mpc_prob} is defined as the sum of all individual subsystem costs:
\begin{multline}\label{eq_glob_cost}
\ell(k)=\sum_{i\in\setS} \ell_i(k)=(\bfx(k) -\bar{\bfx})\trasp Q (\bfx(k)-\bar{\bfx}) \\ + (\bfu(k) - \bar{\bfu})\trasp R (\bfu(k) - \bar{\bfu}),
\end{multline}%
\noindent where $\bfx\triangleq\{x_{i}\}_{i\in\setS}$ and $\bfu\triangleq\{u_{i}\}_{i\in\setS}$ are respectively the global system state and input vectors, and $Q$ and $R$ are block diagonal matrices that gather all local weights. The global state transition matrices employed in~\eqref{eq_mpc_prob_constr} are composed accordingly.
\item Decentralized control (DEC): each subsystem is governed by a local MPC controller based exclusively on the local objective~\eqref{eq_ind_cost}. No explicit coordination between controllers is contemplated.
\end{itemize}
With regard to the way cooperation costs are taken into account, the use of the two indices has been compared in the simulations: in particular, index $(a)$ has been defined for this particular case as $\chi_{12} = (|\setP_1|+|\setP_2|)^2$, i.e., proportional to the square of the number of members of the coalition. Index $(b)$ has been defined as a linear relation on the number of enabled data links $\chi_{12} = n_{l1}+n_{l2}+1$.\par
\begin{figure*}[tb]
  \includegraphics[width=\textwidth,trim = 30mm 20mm 30mm 0mm, clip=true]{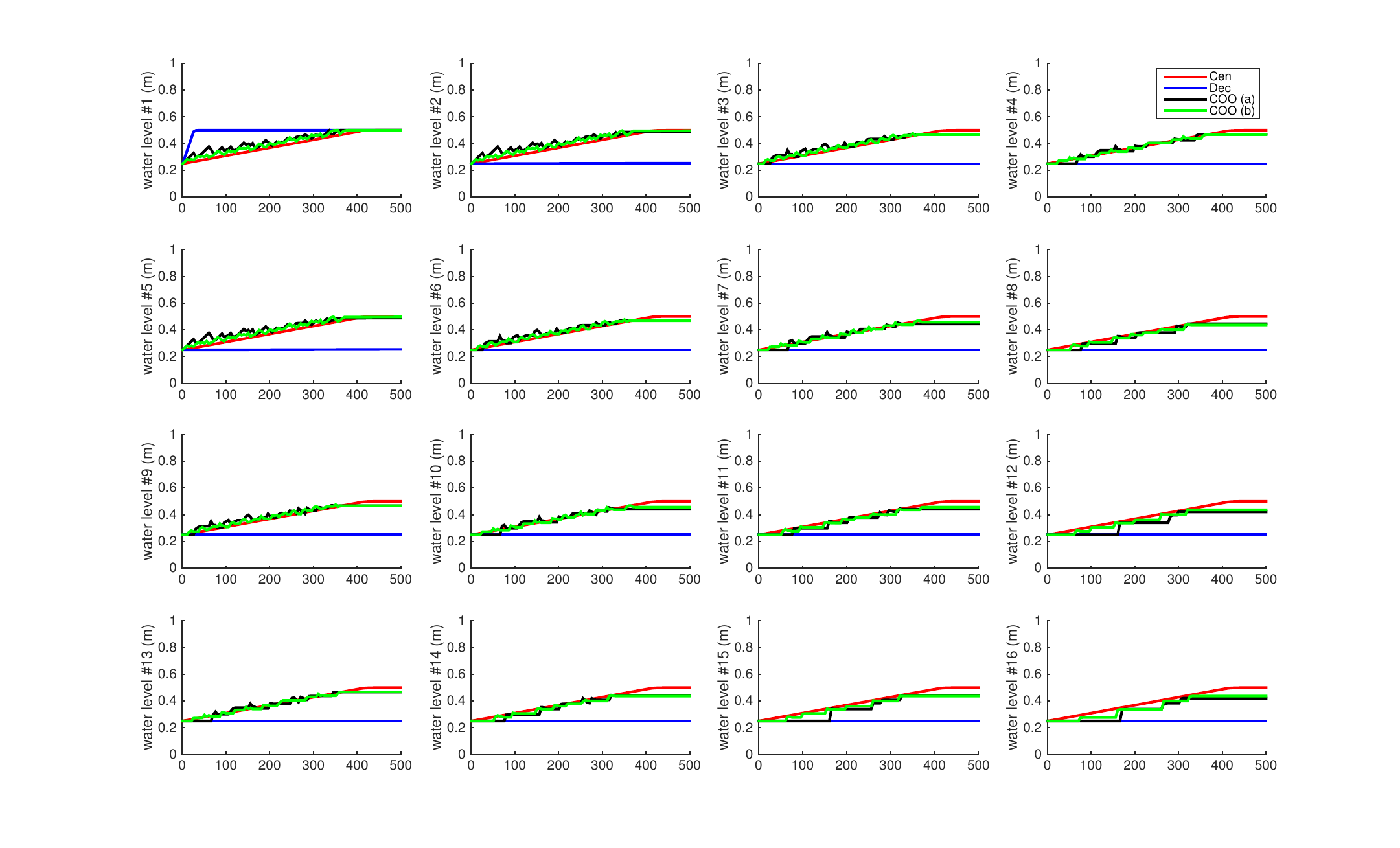}
  \caption{Evolution of the states for the control schemes considered.}\label{fig:waterlevels}
\end{figure*}
A comparison of the results obtained through the autonomous coalition formation dictated by the COO algorithm, with those relative to the CEN and DEC controllers is shown in Fig.~\ref{fig:waterlevels}.\footnote{Plots relative to the CIR criterion coincide with the ones relative to the COO algorithm, and for this reason are not shown in Fig.~\ref{fig:waterlevels}. Indeed, the effect of the application of the CIR criterion is seen on an \emph{a posteriori} cost redistribution.} Note that the results relative to the CEN scheme are not restricted to a central implementation: indeed, similar performances can be attained by distributed MPC architectures (e.g., based on an augmented Lagrangian formulation)~\cite{NegenbornMaestreCSM2014,VenkatThesis2006,Negenborn:07d}. In this case, the absence of communication in the DEC scheme prevents coordination: as a consequence the top-left subsystem is the only one that reaches the target steady state. Notice that the overall results achieved through autonomous coalition formation are in between those relative to CEN and DEC. Indeed, they show a sensible improvement w.r.t. the DEC controller, by actually performing comparably to the CEN controller. This is reflected on the final cumulated cost, reported in Table~\ref{tab_costs} for all the schemes considered in the simulations.\par
\begin{table}[tb]
\centering
	\caption{Accumulated costs relative to simulations in Fig.~\ref{fig:waterlevels}.}
   \begin{tabular}{c c c c c}
     CEN & DEC & COO (a) & COO (b) & \\   \hline
     1.40 & 4.68 & 1.51 & 1.46 &  $\cdot 10^4$
   \end{tabular}
   	\label{tab_costs}
\end{table}
The bottom-up coalitional control scheme promotes the cooperation of groups of subsystems to improve their MPC performance index and, as a consequence, the overall system is driven towards its setpoint. As the error decreases, so decreases the cooperation rate. Indeed, the establishment of any further communication must yield a performance improvement high enough to compensate its costs, modeled through indices $(a)$ and $(b)$. Hence, the active coordination among agents is reduced until the structure of the overall control corresponds to a decentralized architecture once the setpoint is reached. It should be pointed out that the additional penalties imposed in order to constrain the size of coalitions have a price in terms of steady state error. Figure~\ref{fig:waterlevels} shows clearly this effect: the closer a subsystem is located to the external water inflow, the smaller is its final steady state error. As the distance from the source increase, further layers of intermediate links are needed, amplifying the price of the cooperation. This point also raises interesting questions regarding the relationship between the local performance and the position of the subsystems in the network. Besides being a relevant topic in the control engineering literature~\cite{SummersLygeros2014,LiuEtAl2011}, similar problems are under analysis in disciplines such as game theory --- in particular the study of social and economic networks~\cite{Jackson2010,EasleyKleinberg2010}.   

%% file: conclusions.tex
\section{Conclusions}
\label{sec_concl}
Fundamental concepts of game theory constitute a basis for the analysis of the interaction of control agents, as well as for the design of cooperative mechanisms aimed at management of complex systems. Some of these concepts, in particular those developed within the framework of noncooperative games (chiefly the Nash equilibrium) have been extensively applied in the distributed control literature. Notions related with cooperative games, however, are in general less tailored for dynamical environments. As such, their application in control engineering is seldom encountered. Nevertheless, the cooperative games framework is being extended towards the control engineering world thanks to a growing number of pioneering works. A future line of investigation is being defined for the development of control strategies for systems admitting dynamically evolving coalitional structures.\par
The bottom-up approach presented in this paper is a preliminary step towards the application of coalitional control on complex distributed systems (e.g., systems of systems). Ongoing work involves the inclusion of splitting processes into the algorithms, and the extension of the present study to multiple-player cooperative games. The role of cooperation costs on the outcome of the coalition formation, as well as its relation with control optimality, constitutes an interesting topic for future research. A further line of investigation may tackle distributed estimation issues naturally arising in distributed systems.